\documentclass{article}
\usepackage{frascatiphys}
\usepackage{graphicx}
\usepackage[T1, OT1]{fontenc}
\DeclareTextSymbolDefault{\dh}{T1}
\usepackage{hyperref}
\begin{document}
\title{ 
ARE WE REALLY SEEING DARK MATTER SIGNALS FROM THE MILKY WAY CENTER?
}
\author{
Germ\'an A. G\'omez-Vargas        \\
{\em Instituto de Fis\'ica, Pontificia Universidad Cat\'olica de Chile} \\
{\em INFN, Sezione di Roma ``Tor Vergata''}
}
\maketitle
\baselineskip=11.6pt
\begin{abstract}
The center of the Milky Way  is one of the most interesting regions of the $\gamma$-ray sky because of the potential for indirect dark matter (DM) detection. It is also complicated due to the many sources and uncertainties associated with the diffuse $\gamma$-ray emission. Many independent groups have claimed a DM detection in the data collected by the Large Area Telescope on board the Fermi $\gamma$-ray Satellite from the inner Galaxy region at energies below 10 GeV.  However, an exotic signal needs to be disentangled from the data using a model of  known $\gamma$-ray emitters, i.e. a background model. We point out that deep understanding of background ingredients and their main uncertainties is of capital importance to disentangle a dark matter signal from the Galaxy center.
\end{abstract}
\baselineskip=14pt

\section{Introduction}
\begin{figure}[htb]
    \begin{center}
        {\includegraphics[scale=0.44]{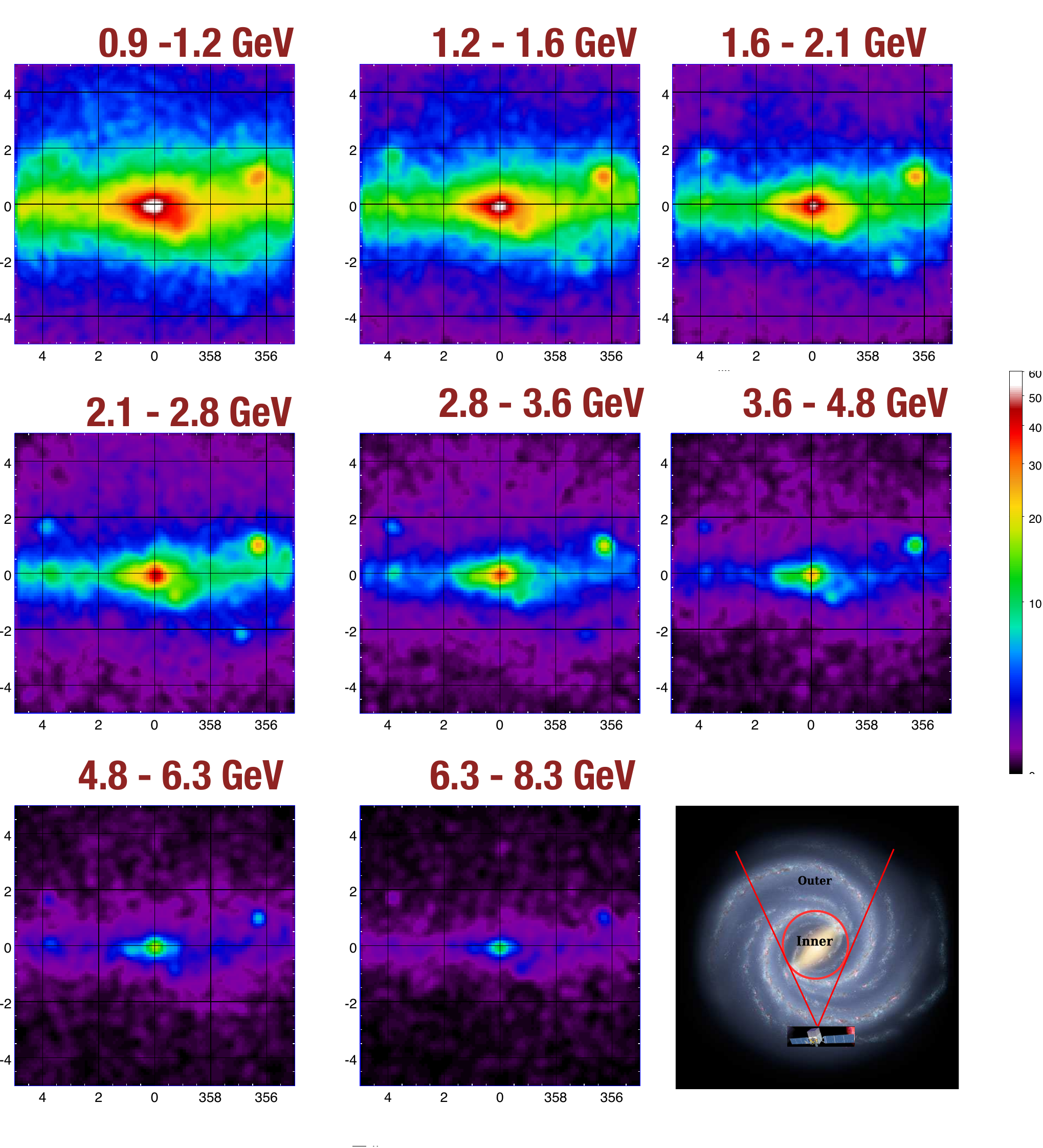}}
        \caption{\it   \textit{Fermi}-LAT view of the Milky Way center in different energy bands (color scale is the same in all maps). In the lower right panel a schematic of the GC view by the \textit{Fermi}-LAT.}
\label{fig:IG}
    \end{center}
\end{figure}

The Fermi Large Area Telescope (\textit{Fermi}-LAT), the main instrument of the Fermi satellite, which has been in orbit since June $11$, $2008$~\cite{REF:2009.LATPaper},  performs $\gamma$-ray measurements covering an energy range from $\sim 20$ MeV to $>300$ GeV over the whole celestial sphere. The  \textit{Fermi}-LAT has detected point and small extended sources, e.g. blazars, supernova remnants (SNRs) and pulsars~\cite{2fgl}, and a strong diffuse component in the whole sky first observed by the OSO-3 satellite in the inner Galaxy region~\cite{OSO}. See images of the region around the Galaxy center at different energies as seen by  \textit{Fermi}-LAT in figure~\ref{fig:IG}. The main contribution of the emission detected in the direction of the inner Galaxy is  made of: outer Galaxy, true inner Galaxy, foreground emission, unresolved sources, point or small extended sources, extragalactic emission, possible dark matter (DM) contribution, and cosmic ray (CR) instrumental background; see lower right panel of figure \ref{fig:IG}. 
%

The possible DM contribution to $\gamma$-ray data in the inner Galaxy direction can be calculated convolving some fundamental characteristics of DM candidates with the distribution of DM, $\rho$, as predicted by cosmological N-body simulations. The basic characteristics of DM candidates  relevant for $\gamma$-ray calculations are: DM mass $m_{DM}$, thermal average of DM-Standard Model (SM) cross section times DM relative velocity $\langle\sigma_{ann} v\rangle$  and the number of $\gamma$ rays produced per annihilation $N_{\gamma}$\footnote{assuming that DM particle is stable}. In this way we can calculate the flux of DM-induced $\gamma$ rays:

\begin{eqnarray}
\frac{d\Phi_{\gamma}}{dE}(E) = \frac{\langle\sigma_{ann}v\rangle}{8\pi m^2_{\chi}}\sum{Br^i \frac{dN^i_{\gamma}}{dE}(E)} \int_{\Delta\Omega}{d\Omega\int{d\lambda} \rho^2(\lambda,\Psi)}.
\end{eqnarray}

The DM density $\rho$ integrated over the line of sight $\lambda$ and the angular region of the sky $\Delta\Omega$ is the so called {\it J-factor}. Assuming that DM is made of weakly interacting massive particles (WIMPs) produced thermally in the early universe, the value of $\langle\sigma_{ann} v\rangle$ must be $\approx 3\times10^{-26}$ cm$^3$/s to produce observed DM relic abundance. The largest value of the J-factor is in the Galactic Center, where due to the large uncertainty in $\rho$, DM can either, overshoot data or only contribute modestly to the observed emission. Using the former possibility the thermal cross section was excluded for a large range of $m_{DM}$~\cite{GomezVargas:2011ph}. In the latter case, DM-induced $\gamma$ rays would appear as an exotic contribution in {\it Fermi}-LAT data of the region around the GC. We need to understand the non-exotic contributions, i.e. the background, in order to disentangle a possible DM signal.

Many independent groups have claimed a DM detection in the data collected by the Fermi-LAT from the GC region~\cite{2009arXiv0912.3828V,Hooper:2010mq,2010arXiv1012.2292M,Hooper:2011ti,Abazajian:2012pn,Macias:2013vya,Macias:2013vya,Abazajian:2014fta,Daylan:2014rsa,Calore:2014xka}. This source may be due to DM particles annihilating, but other plausible phenomena may be responsible for this. All these analyses are based on the subtraction of background models: diffuse interstellar emission and point sources~\cite{2fgl}, from the data. To build these diffuse models there are two approaches, to use CR propagation codes such as GALPROP\footnote{For a detailed description of the GALPROP code and the most recent release that we use in this work (version 54), we refer the reader to the dedicated website \texttt{http://galprop.stanford.edu}}, or the template fitting method. The ingredients needed for both approaches and their uncertainties are presented in section 2. The main idea of the methods and their issues are discussed in section 3. Section 4 we address the question: are we seeing DM signals from the Galaxy center?

\section{Ingredients for building diffuse models and their uncertainties}

\begin{figure}[!tb]
    \begin{center}
  \includegraphics[scale=0.4]{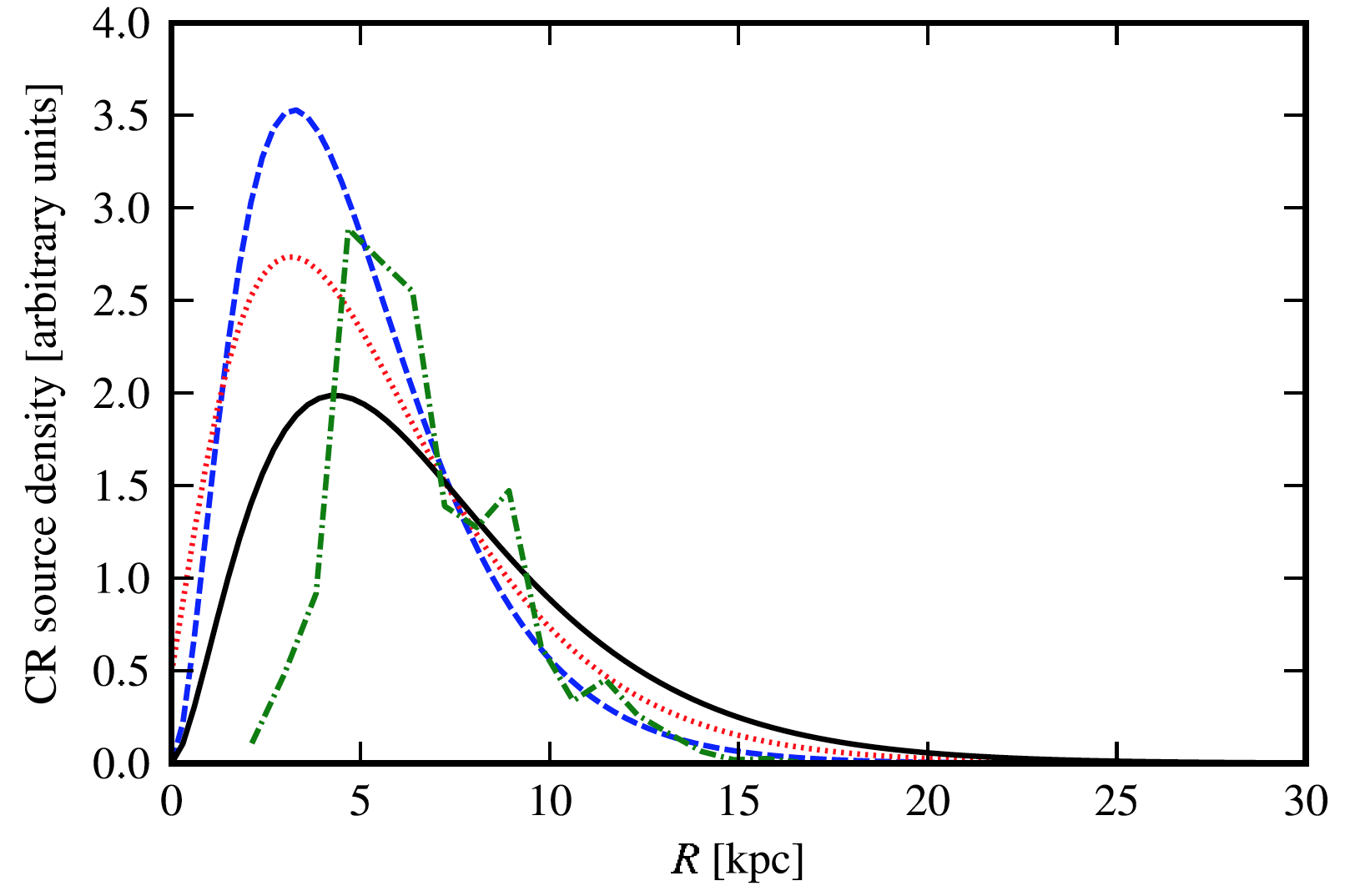}
\caption{Primary CR nuclei and electron source distribution for the large-scale diffuse Galactic models used in \cite{diffuse2}. Solid black, SNR (Lorimer). Dashed blue, pulsars (Lorimer). Dotted red, pulsar (Yusifov). Dash-dotted green, OB-stars} 
\label{fig:CR_dis}
    \end{center}
\end{figure}

The $\gamma$-ray diffuse emission in the inner Galaxy is created by interaction of CR with interestellar gas (pion decay and bremsstrahlung), radiation fields (ICS), and  magnetic fields (synchrotron).  So, to build diffuse emission models of this region some basic ingredients are needed, they are listed here:

\begin{itemize}

\item{\textbf{Molecular Hydrogen H2:}}
Concentrated mostly in the plane. The main tracer is CO. Distance information from velocity and a rotation curve is used to assign the gas to galactocentric rings. The standard method of assigning velocity to distance, in order to create the rings, breaks down toward the GC. The so call Xco factor to convert CO to H2 column density is believed to vary as a function of the galactocentric radius. However, the exact form of the variation is not well know.

\item{\textbf{Atomic Hydrogen HI:}}
The 21 cm line HI map used is from \cite{Kalberla:2005ts}. As for H2, distance information from velocity and a rotation curve is used to assign the gas to galactocentric rings. The main uncertainty comes from the spin temperature $T_s$. The code uses a single $T_s$ value among many possibilities. Indeed, HI is a mixture of various phases, observations of $T_s$ show it to vary from tens of K up to thousands of K, so that the adoption of a single $T_s$\footnote{In \cite{diffuse2} only 2  $T_s$ extreme values were used, 150 }  is in any case an approximation.

\item{\textbf{Galactocentric rings toward the GC:}} \label{grings}
The kinematic resolution of the method used to relate velocity and distance vanishes for directions near the GC. We linearly interpolate each annulus independently across the range $|l|<10^{\circ}$ to get an estimate of the radial profile of the gas. Nevertheless, the innermost annulus is entirely enclosed within the interpolated region, necessitating a different method to estimate its column density. For HI the innermost annulus contains $\sim$60\% more gas than its neighbouring annulus. This is a conservative number. For CO, we assign all high velocity emission in the innermost annulus. See Appendix 2 of \cite{diffuse2} for more details.

\item{\textbf{Interstellar Radiation Field (ISRF):}}
Emission from stars, and the scattering, absorption, and re-emission of absorbed starlight by dust in the ISM. The \texttt{FRaNKIE} code \cite{frankie} is used to model the distribution of optical and infrared (IR) photons throughout the Galaxy. Further details about the ISRF model and recent developments about modelling this component, can be found in Appendix 3 of  \cite{diffuse2}. The main uncertainty is the overall input stellar luminosity and how it is distributed amongst the components of the model (bulge, thin and thick disk, and halo)

\item{\textbf{CR injection and propagation:}}
SNRs are widely accepted as the main sources of CRs. However, their distribution is not well determined. Pulsars are SN explosion end states and their distribution is better determined than SNRs, but still, it suffers from observational biases. CR propagation is not well understood and its uncertainties involve spectra injection, transport parameters, halo size, etc. In figure \ref{fig:CR_dis} we present the distribution of  CR widely used, as e.g. in \cite{diffuse2}.

\item{\textbf{Inverse-Compton Scattering (ICS):}}
Optical photons are the principal target for high energy electrons to produce ICS emission in the energy range $\sim 50$ MeV -100 GeV. The ICS template is brightest
in the direction of the inner Galaxy, and while it should be smooth because of the physics of radiation in the Galaxy, there are most likely fluctuations in that component that are not
modelled with GALPROP. A dedicated study of the ISRF and the CR source distribution in the direction of the inner Galaxy to be able to estimate this contribution is needed.

\end{itemize}

There are some extended sources not included here such as the Fermi Bubbles \cite{Dobler:2009xz,Su:2010qj,Fermi-LAT:2014sfa} and Loop I \cite{Casandjian} . Templates that model these sources must be included in order to have an accurate description of the $\gamma$-ray sky observed by the \textit{Fermi}-LAT.

\section{Recipes}

\begin{figure}[!tb]
    \begin{center}
        {\includegraphics[scale=0.35]{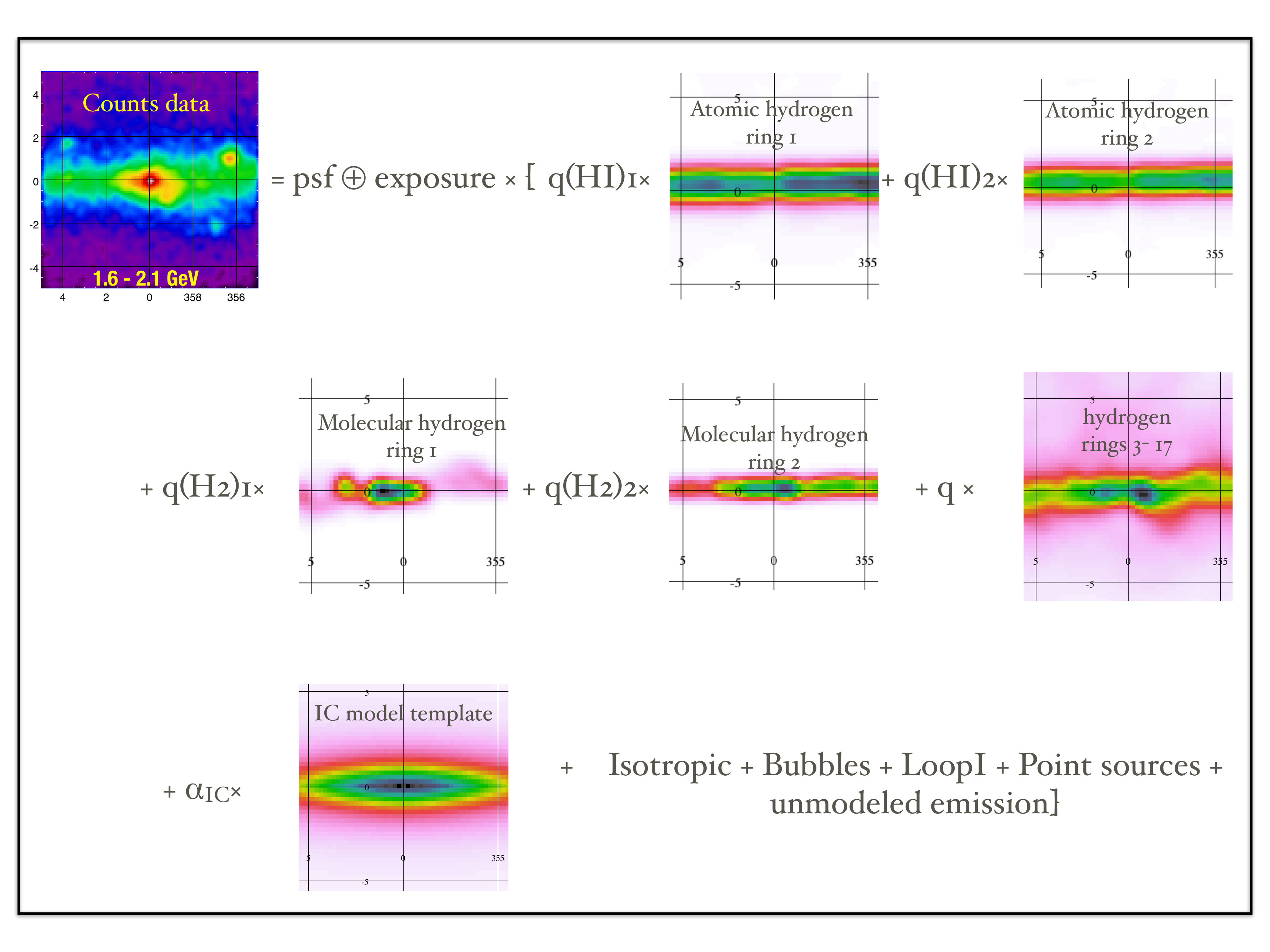}}
        \caption{\it  Template fitting method. At different energy bins, templates  correlated with gas, IC and some extended sources are directly fitted to the data. }
\label{fig:IG_templates}
    \end{center}
\end{figure}

\begin{itemize}
\item{\textbf{CR propagation codes:}} GALPROP code calculates the propagation of CR, and computes diffuse $\gamma$-ray emission in the same framework. Each run using specific realistic astrophysical inputs together with theoretical models corresponds to a potentially different background model for DM searches. By varying these inputs within their limits, many diffuse emission models can be created. GALPROP accounts for effects such as diffusion, reacceleration, and energy loss via mechanisms such as synchrotron radiation.  In \cite{diffuse2} different GALPROP models were compared with data, finding that all of them are in good (~20\%) agreement with all sky data

\item{\textbf{Template fitting method:}} At some particular energy, the $\gamma$-ray intensity is modelled as a linear combination of gas column-density map template, a predicted IC intensity map and a residual intensity of unmodeled emission. Figure \ref{fig:IG_templates} presents the idea of this method. The diffuse models provided by the {\it Fermi}-LAT collaboration\footnote{http://fermi.gsfc.nasa.gov/ssc/data/access/lat/BackgroundModels.html} to study point or small extended sources are created using this method.

\end{itemize}

All the template-based models provided by the {\it Fermi}-LAT collaboration are fitted to the whole-sky with the purpose of serving as background models for analysis of pointlike or small sources, and as such tried to pick up as much extended emission as possible. The double fit (original one plus GC fit) introduces complications in the interpretation of the results which are not trivial to understand.

Any model based on the gas maps created for full sky analysis will not be very good in the inner Galaxy by design. The linear interpolation used for the distance estimator is a very basic approximation and can not be used to estimate the diffuse emission in that inner Galaxy region. A very dedicated study on the gas templates is needed to understand that region.

The most important message is that if one wants to study extended emission in the direction of inner Galaxy region there is no ready-made solution in terms of a diffuse background model to use. None of the models up to now are adequately describing $\gamma$-ray emission from that region.

\section{Are we really seeing DM signals from the Milky Way center?}

Maybe yes, it is clear from maps in figure~\ref{fig:IG} that there is an extended $\gamma$-ray source in the very GC, whatever its nature is. But, we can not be sure it is a DM signal as far as we do not understand the background at the level needed to characterise its spectrum and morphology. Beside this, there is an implicit assumption in the modelling discussed above: steady state. It is a strong assumption since it is very likely that the GC  has a violent history, two recent papers \cite{Gonzalez-Morales:2014eaa,Petrovic:2014uda} present cases where past activity in the GC may yield $\gamma$-ray emission with similar properties to DM sources.

We need new molecular and atomic gas, CR and $\gamma$-ray data to shed light on the nature of the GC region at high energies. We already have new gas data waiting to be studied in the context of $\gamma$-ray astronomy \cite{australia,3d,Plank}. Regarding new $\gamma$-ray data, from  space there are some proposals to build new satellites \cite{gammalight,CALET,gamma400,astromev}. From Earth, the Cherenkov Array Telescope (CTA) will provide insight on the mysterious phenomena at the Milky Way center \cite{CTA}.

\section{Acknowledgements}
Thanks to Gu{\dh}laugur J\'ohannesson and Luigi Tibaldo. Also thanks to Luis Labarga and Carlos Mu\~noz for comments on the manuscript. The work of GAGV was supported by Conicyt Anillo grant ACT1102. GAGV thank for the support of the Spanish MINECO's Consolider-Ingenio 2010 Programme under grant MultiDark CSD2009-00064. This work was also supported in part by MINECO under grant FPA2012-34694. 
\end{document}